# Magnetoelectric-field microwave antennas: Far-field orbital angular momenta from chiral-topology near fields

M. Berezin, E. O. Kamenetskii, and R. Shavit

Microwave Magnetic Laboratory,
Department of Electrical and Computer Engineering,
Ben Gurion University of the Negev, Beer Sheva, Israel

December 3, 2015

**Abstract**
The near fields in the proximity of a small ferrite particle with magnetic-dipolar-mode (MDM) oscillations have space and time symmetry breakings. Such MDM-originated fields – called magnetoelectric (ME) fields – carry both spin and orbital angular momentums. By virtue of unique topology, ME fields are strongly different from free-space electromagnetic (EM) fields. In this paper, we show that because of chiral topology of ME fields in a near-field region, far-field orbital angular momenta (OAM) can be observed, both numerically and experimentally. In a single-element antenna, we obtain a radiation pattern with an angular squint. We reveal that in far-field microwave radiation a crucial role is played by the ME energy distribution in the near-field region.

## I. INTRODUCTION

It is well known that EM fields can carry not only energy but also angular momentum. The angular momentum is composed of spin angular momentum (SAM) and orbital angular momentum (OAM) describing its polarization state and the phase structure distribution, respectively. The research on OAM of EM fields was not attractive until Allen *et al.* investigated the mechanism of OAM in laser modes [1]. Henceforth, more and more attention has been paid to OAM in both optical- and radio-wave domains. In contrast to SAM, which has only two possible states of left-handed and right-handed circular polarizations, the theoretical states of OAM are unlimited owing to its unique characteristics of spiral flow of propagating EM waves [2]. Therefore, OAM has the potential to tremendously increase the spectral efficiency and capacity of communication systems [3]. Among numerous investigations in OAM effects, one of the subjects of intensive recent studies in optics concerns relations between near-field chirality and far-field OAM. For different types of chiral polaritonic lenses, it was shown that near-field chirality can lead to tailoring optical OAM in the far-field region [4 – 9]. These results offer new opportunities for far-field 3D light shaping with distinct handedness.

While numerous experiments on OAM were realized in optical frequencies, the concept of OAM in microwave frequencies is relatively novel. It was shown [10 – 12] that generating OAM in microwave frequencies can be realized based on circular antenna arrays. In these antenna structures, microwave radiation elements are situated at distances about a half of the electromagnetic (EM) wavelength. It is worth noting, however, that no chiral polaritonic lenses have been proposed in microwaves till now. Moreover, no concepts on the microwave near-field chirality that can lead to tailoring microwave OAM in the far-field, have been suggested.

The way to realize chiral polaritonic lenses in microwaves is to find subwavelength particles which exhibit effective resonant interactions with microwave fields. Since resonance frequencies of plasmon- and exciton-polariton states are very far from microwave frequencies, the main ideas and results of the optical subwavelength photonics cannot be used in microwaves. In microwaves, however, there exist subwavelength particles (with sizes much less than the free-space electromagnetic wavelength $\lambda_0$), which are distinguished by specific magnetic-dipolar-mode (MDM) oscillations [13]. In recent studies it was shown that MDM oscillations in a quasi-2D ferrite disk are macroscopically coherent quantum states, which experience broken mirror symmetry and also broken time-reversal symmetry. One observes very effective resonant interactions of ferrite-disk particles with microwave fields resulting in appearance of MDM-polariton states. Free-space microwave fields, emerging from magnetization dynamics in quasi-2D ferrite disk, carry spin and orbital angular momentums and are characterized by power-flow vortices and non-zero helicity. Symmetry properties of these fields – called magnetoelectric (ME) fields – are different from symmetry properties of free-space electromagnetic (EM) fields. For an incident electromagnetic field, the MDM ferrite disk looks as a trap with focusing to a ring, rather than a point [14 – 21].

In this paper, we show that due to the chirality and the OAM of near fields originated from MDM ferrite particles, the nontrivial far-field OAM can be generated in microwaves. The topological near-field regions with broken symmetry we call chiral MDM-polariton lenses. We investigate a novel microwave antenna with a MDM ferrite disk as a basic building block for controlling the far-field electromagnetic radiation. The antenna has symmetrical geometry. At the MDM resonance, topological singularities of the ME near fields cause chiral electric current distributions on metal surfaces of the antenna. This results in the topological singularities in the far-region radiation fields. The MDM antenna continuously modulates both amplitude and phase in the diffraction field to shape twisted radiation pattern. This forms a far-field radiation pattern with a strong and controllable squint. Changing the input/output ports and a direction of a bias magnetic field allows easy manipulation of spatial degrees of freedom of microwave photon states. Such an effect of a spatial mode division by a single radiation element is unique from a fundamental point of view. This effect can be attractive for development of novel microwave radiation systems with controllable phase structure distribution.

The paper is organized as follows. In Section II, we describe the MDM-resonance antenna. To find a transfer of the topological ME effects in the far-field region, this microwave antenna has the MDM-resonance structure as the only resonant element. In Section III, we demonstrate, both numerically and experimentally, the effect of tailoring ME near-field chirality to a microwave OAM in the far-field region. This effect is verified by the presence of angular squints in radiation patterns. Section IV presents a theoretical insight into the MDM-resonance far-field orbital angular momenta. We analyze the ME-field helicity conservation law. We show that near a ferrite disk at the MDM-resonance frequencies one has the regions with positive helicity and negative helicity. For the entire space volume, we have the "helicity neutrality". We analyze numerically the field topology in radiation near- and far-field regions. We show that the far-field orbital angular momenta, observed at the MDM-resonance frequencies, appear due to the near-field topology of the "helicity dipoles" ("ME dipoles"). In Section V, we conclude our studies.

## II. MDM ANTENNA: STRUCTURE AND MICROWAVE CHARACTERISTICS

To find the transfer of the MDM-resonance topological ME effects [14 – 21] in the far-field region, a microwave antenna should not have other resonant elements except the MDM resonant structure. For this purpose, we use a rectangular waveguide with a hole in a wide wall and the diameter of this hole is much less than a half wavelength of microwave radiation. At the MDM resonance, the topological singularities in the far-region radiation fields appear due to topological singularities of the ME near fields which, in their turn, cause chiral electric current distributions on the external surface of a waveguide wall. Such current singularities are well distinguished in a microwave antenna with symmetrical geometry. For this reason, the hole in a waveguide wide wall is situated symmetrically. For the numerical and experimental studies we use a microwave antenna shown in Fig. 1. This is a waveguide radiation structure with a hole in a wide wall and a thin-film ferrite disk as a basic building block. A hole in a wide wall has a diameter of 8 mm, which is much less than a half wavelength of microwave radiation at the frequency regions from 8.0 GHz till 8.5 GHz, used in our studies. The yttrium iron garnet (YIG) disk is placed inside a $TE_{10}$-mode rectangular X-band waveguide symmetrically to its walls so that a disk axis is perpendicular to a wide wall of a waveguide. The disk diameter is 3 mm and the thickness is 0.05 mm. The ferrite disk is normally magnetized by a bias magnetic field $H_0 = 4760$ Oe; the saturation magnetization of a ferrite is $4\pi M_s = 1880$ G. For better understanding the field structures, in a numerical analysis we consider a ferrite disk with a very small linewidth of $\Delta H = 0.1$ Oe. In experimental studies, a ferrite-disk sample has the pointed above parameters except a linewidth which is equal to $\Delta H = 0.6$ Oe.

Fig. 2 shows the numerical reflection spectral characteristic for the $TE_{10}$ waveguide mode in our microwave antenna. Because of the presence of a radiation hole, the spectrum is quite emasculated compared to a multiresonant spectrum of MDM oscillations observed in non-radiating microwave structures [16 – 19]. The peaks 1-1′ and 3-3′ correspond to the radial MDMs, while the peak 2-2′ is the azimuth-mode magnetic-dipolar resonance [22]. The forms of the resonance peaks in Fig. 2 give evidence for the Fano-type interaction [19, 23, 24]. The underlying physics of the Fano resonances finds its origin in wave interference which occurs in the systems characterized by discrete energy states that interact with the continuum spectrum. In our structure, the discrete states are due to MDM resonances and an entire continuum is composed by the internal waveguide and external free-space regions. The forward and backward propagating modes within the waveguide are coupled via the defects. This coupling becomes highly sensitive to the resonant properties of the defect states. For such a case, the coupling can be associated with the Fano resonances. In the corresponding transmission dependencies, the interference effect leads to either perfect transmission or perfect reflection, producing a sharp asymmetric response. We have a "bright" and a "dark" resonances, which produce the Fano-resonance form in the reflection spectra. In our study we will use the first "bright" peak – the peak 1′ – where the most intensive radiation is observed. At the frequency of the resonance peak 1′, the field structure near a ferrite disk is typical for the main MDM excited in a closed waveguide system [16 – 19]: there are rotating electric and magnetic fields, the active power flow vortex, and non-zero field helicity.

Fig. 3 shows the experimental transmission spectral characteristic. Experimentally, we observe only the radial MDMs: the resonance peaks 1-1′ and 3-3′. Moreover, because of certain geometrical non-symmetry in an experimental structure, the high-order radial peak 3-3′ in Fig. 3 is excited stronger than such a peak in a numerical characteristic (see Fig. 2).

Because of chiral topology of the MDM-resonance near fields [14 – 21], far-field orbital angular momenta can be observed both numerically and experimentally. This effect of tailoring the near-field chirality to a microwave OAM in the far-field region can be well verified by appearance of angular squints in radiation patterns. Usually, angular squints in microwave radiation patterns are obtained due to the free-space interference processes with use of an array of radiation elements [10 – 12]. In our single-element structure, we obtain radiation patterns with the angular squints due to specific topology of the ME near fields.

For an analysis of radiation patterns we use a spherical coordinate system, shown in Fig. 4. Fig. 5 shows numerically calculated far-field radiation patterns (directivity) for two cut planes, $zx$ plane ($\varphi = 0°$) and $zy$ plane ($\varphi = 90°$), at the nonresonance frequency 8.125GHz. The wave in a waveguide propagates from port 1 to port 2. In this case, we have a squint in the $zy$ plane, along a negative direction of $y$ axis. When wave propagates from port 2 to port 1, the squint is along a positive direction of $y$ axis. The observed squints in the $zy$ plane are regular squints which take place due to a small phase delay for the wave when it propagates in a region a waveguide hole. There are no squints in the $zx$ plane and the radiation patterns are the same for two opposite directions of a bias magnetic field.

At the frequency of the MDM resonance, squints in the $zx$ plane occur together with the existing regular squints in the $zy$ plane. The squint in $zy$ plane is the same as in Fig. 5. As a result, the angular squints are observed in numerically calculated far-field radiation patterns. Such angular squints are shown in Figs. 6 – 8. Fig. 6 shows single-element 2D and 3D radiation patterns for two cut planes, $\varphi = 0°$ and $\varphi = 90°$, at the MDM-resonance frequency 8.139GHz. The wave in a waveguide propagates from port 1 to port 2. Fig. 7 shows 2D radiation patterns in the $zx$ plane ($\varphi = 0°$) at the MDM-resonance frequency 8.139GHz and two opposite directions of the wave propagation in a waveguide. Fig. 8 shows the resonance-frequency radiation patterns in the planes $\varphi = 0°$ and $\varphi = 90°$ for two opposite directions of the bias magnetic field and the same direction of the wave propagation in a waveguide (from port 1 to port 2).

The numerical results are well verified by our experimental studies. Fig. 9 shows the numerically simulated and experimentally measured radiation patterns for a cut plane $\varphi = 0°$ at the MDM-resonance frequency 8.139GHz. Wave propagates from port 1 to port 2 and a bias magnetic field is directed upward. The measurements were done in a sector with variation of a spherical coordinate from $\theta = -120°$ to $\theta = 120°$, where the angular squints are expected. Experimental radiation patterns in Fig. 10 well illustrates a role of the MDM resonance in obtaining angular squints. Fig. 10 (*a*) shows the measured normalized radiation patterns for nonresonant (8.124GHz) and resonant (8.132GHz) frequencies. Figs. 10 (*b*) and 10 (*c*) present enlarged pictures of these radiation patterns.

### III. MDM-RESONANCE FAR-FIELD ORBITAL ANGULAR MOMENTA: A THEORETICAL INSIGHT

The shown angular squints in far-field radiation patterns is observed because of the ME-field helicity and ME-energy distributions in the radiation near-field area. At the MDM-resonance frequencies, near a ferrite disk one has the regions with positive helicity (positive ME energy) and negative helicity (negative ME energy). These regions constitute the "helicity dipoles" ("ME dipoles"). We will show that the far-field orbital angular momenta, observed at the MDM-resonance frequencies, appear due to the near-field topology of these "ME dipoles".

**A. ME-field helicity conservation law and "ME dipoles"**

At the MDM resonances ($\omega = \omega_{MDM}$), for the ME-field structure in a vacuum region near a ferrite disk, the magnetic field is a potential field, $\vec{H} = -\vec{\nabla}\psi$, while the electric field has two parts: the curl-field component $\vec{E}_c$ and the potential-field component $\vec{E}_p$. The curl electric field $\vec{E}_c$ in vacuum is defined from the Maxwell equation $\vec{\nabla} \times \vec{E}_c = -\mu_0 \frac{\partial \vec{H}}{\partial t}$. The potential electric field $\vec{E}_p$ in vacuum is calculated by integration over the ferrite-disk region, where the sources (magnetic currents, $\vec{j}^{(m)} = \frac{\partial \vec{m}}{\partial t}$) are given. Here $\vec{m}$ is dynamical magnetization in a ferrite disk. Numerical studies show that for the ME near fields, we have $|\vec{E}_p| \gg |\vec{E}_c|$. For time-harmonic fields ($\propto e^{i\omega t}$), and with representation of the potential electric field as $\vec{E}_p = -\vec{\nabla}\phi$, the time-averaged helicity density parameter in a vacuum near-field region is calculated as [17 – 19, 21]:

$$F = \frac{\omega_{MDM}\varepsilon_0}{4} \operatorname{Re}(\vec{E} \cdot \vec{B}^*) = -\frac{\omega_{MDM}\varepsilon_0}{4} \operatorname{Re}[\vec{\nabla}\phi \cdot \vec{B}^*] = -\frac{\omega_{MDM}\varepsilon_0}{4} \operatorname{Re}[\vec{\nabla} \cdot (\phi\vec{B}^*)]. \quad (1)$$

Here we took into account that $\vec{\nabla} \cdot \vec{B} = 0$. We can also introduce a quantity of the time-averaged ME-energy density [21]:

$$\operatorname{Re}[\vec{\nabla} \cdot (\phi\vec{B}^*)] \equiv -\eta \quad (2)$$

and represent the helicity density $F$ as

$$F = \frac{\omega_{MDM}\varepsilon_0}{4} \eta. \quad (3)$$

An integral of the ME-field helicity density (or ME-energy density) over the entire near-field vacuum region of volume $V$ (which excludes a region of a ferrite disk) we define as the helicity $\mathcal{H}$:

$$\mathcal{H} \equiv \int_V F dv = \frac{\omega_{MDM}\varepsilon_0}{4} \int_V \eta dV = -\frac{\omega_{MDM}\varepsilon_0}{4} \operatorname{Re}\int_V [\vec{\nabla} \cdot (\phi\vec{B}^*)] dV = -\frac{\omega_{MDM}\varepsilon_0}{4} \operatorname{Re}\oint_S [\phi\vec{B}^* \cdot \vec{n}] dS. \quad (4)$$

From this integral relation, we can conclude that when the normal component of $\vec{B}$ vanishes at some boundary inside which the fields are confined (i.e. when $\vec{B} \cdot \vec{n} = 0$ at the boundary), the quantity $\mathcal{H}$ is equal to zero. The quantity $\mathcal{H}$ is also equal to zero when the fields are with finite energy and the quantity $\phi\vec{B}$ decreases sufficiently fast at infinity.

Inside the vacuum-region volume $V$ there could be a region $k$ ($1 \leq k \leq N$) with volume $V_k^{(+)}$ where the helicity is a non-zero positive quantity:

$$\mathcal{H}_k^{(+)} = \frac{\omega_{MDM}\varepsilon_0}{4} \int_{V_k^{(+)}} \eta_k^{(+)} dV > 0 \quad (5)$$

and a region $l$ ($1 \leq l \leq N$) with volume $V_l^{(-)}$ where the helicity is a non-zero negative quantity:

$$\mathcal{H}_l^{(-)} = \frac{\omega_{MDM}\varepsilon_0}{4} \int_{V_l^{(-)}} \eta_l^{(-)} dV < 0. \quad (6)$$

For the entire volume $V = \sum_{k=1}^{N} V_k^{(+)} + \sum_{l=1}^{N} V_l^{(-)}$, we have the "helicity neutrality":

$$\mathcal{H} = \sum_{k=1}^{N} \mathcal{H}_k^{(+)} + \sum_{l=1}^{N} \mathcal{H}_l^{(-)} = 0. \qquad (7)$$

The regions with positive and negative helicities are the regions with positive and negative ME energies, respectively. The time-averaged ME-energy density $\eta^{(+)}$ can be considered also as the time-averaged density of the positive "helicity charge" while the quantity $\eta^{(-)}$ – the time-averaged density of the negative "helicity charge". Because of the "helicity neutrality", the "helicity dipoles" ("ME dipoles") may occur in a vacuum region near a MDM ferrite disk. More generally, the "helicity multipoles" ("ME multipoles") can exist. As we will show in the present studies, just because of the near-field in-plane topology of the "helicity dipoles" ("ME dipoles") the far-field orbital angular momenta can be observed at the MDM-resonance frequency.

The analyzed above helicity parameter $F$ can be normalized with respect to the field amplitudes. With such a normalization, one have definite information on the time-averaged angle between the electric and magnetic fields. The normalized helicity parameter is expressed as

$$\cos\alpha = \frac{\mathrm{Im}\left\{\vec{E}\cdot\left(\vec{\nabla}\times\vec{E}\right)^*\right\}}{|\vec{E}||\vec{\nabla}\times\vec{E}|} = \frac{\mathrm{Re}(\vec{E}\cdot\vec{H}^*)}{|\vec{E}||\vec{H}|}. \qquad (8)$$

Certainly, for a regular electromagnetic field $\cos\alpha = 0$. For positive helicity (the positive "helicity charge"), the time-averaged angle between the electric and magnetic fields is less then $90°$ ($\cos\alpha < 0$). For negative helicity (the negative "helicity charge"), the time-averaged angle between the electric and magnetic fields is more then $90°$ ($\cos\alpha > 0$).

**B. Field topology in radiation near- and far-field regions**

The near-field structure of our antenna can be conventionally divided in two sub-regions. The first one shows the fields in close proximity of a disk and the second one describes the fields near a radiation hole. In close proximity of a ferrite disk (on scales about tens micrometers), the MDM-resonance field structure is not different from such a structure in a closed (non-radiating) microwave waveguide with an embedded ferrite disk. This field structure are described in details in Refs. [17, 18, 21]. At the same time, near a hole of a radiating microwave waveguide, the MDM-resonance field structure exhibits very specific topological properties which are strongly dependable on external parameters of a system. These properties of the fields near a radiation hole are illustrated in Figs. 11 – 17.

Fig. 11 shows the Poynting-vector distributions on a vacuum plane near a radiation hole for different combinations of two external parameters: a direction of a bias magnetic field and a direction of the power flow in a waveguide. A vacuum plane is placed below a radiation hole and perpendicular to a disk axis. For the same vacuum plane below radiation hole, Fig. 12 shows the normalized helicity parameter, $\cos\alpha$. Figs. 11 and 12 clearly illustrate that non-symmetries in power-flow vortices are in definite correlation with the helicity properties of ME-field areas on the $xy$ vacuum plane. In Fig. 12, the arrows in right-upper corners show directions of "ME dipoles". There are directions from regions of positive ME helicity to regions of negative ME helicity.

The correlation of the near-field topological structure with the helicity-dipole orientation (the ME-dipole orientation) appears also in a vacuum region above a radiation hole. Fig. 13 shows the ME-field areas on the cross-section *zy* plane in the near-field region. One can see that the ME-field areas (originated from a MDM ferrite disk) significantly "get out" through a radiation hole. Fig. 14 shows the ME-field areas on the *xy* vacuum plane above a radiation hole. Similar to Fig. 12, in Fig. 14, the arrows in right-upper corners show directions of "ME dipoles". There are directions from regions of positive ME helicity to regions of negative ME helicity. One can see that "ME dipoles" above a radiation hole are rotated by 90 degrees with respect to "ME dipoles" below a radiation hole. This rotation can be or clockwise, or counterclockwise dependending on direction of a bias magnetic field. Fig. 15 shows Poynting-vector distributions in a vacuum cylinder above a hole. In this figure, the regions *A* and *B* conventionally designate the positive (divergence of vectors) and negative (convergence of vectors) topological charges in the power-flow-density distributions. When we trace the lines from regions *A* to regions *B*, we see that these lines are perpendicular to the directions of ME dipoles.

ME-field topology strongly influences on distributions of the electric fields and currents on an external metal surface of a waveguide. Figs. 16 and 17 show such distributions at the MDM resonance frequency. The regions *C* and *D* designate the positive and negative surface electric charges on a surface of a hole. The observed chiral forms of a front of an electric field and a surface electric current determine chiral-topology of the fields radiating by a MDM microwave antenna. The far-field structure is strongly determined by ME-field topology of the near fields. This is illustrated by Figs. 18 and 19. Fig. 18 shows intensity of an electric field on the *xy* vacuum plane situated at distance about $3\lambda_0$ above a radiation hole. In this figure, we traced projections on this *xy* plane the *AB* lines which connect the regions of positive and negative topological charges of the power-flow-density distributions shown in Fig. 15. We also showed, by small stars, where on the *AB* lines we have maximal intensities of the electric field. One can see how the direction of a squint in the far-field radiation pattern is correlated with the non-symmetry of the power-flow-density distribution in a near-field region. There is an evident correlation between the squint positions and the directions and orientations of the *AB*-line vectors. The same small stars (where on *AB* lines we have a maximal intensity of an electric field) on the same *xy* plane are shown in Fig. 19 in correlation with the three vectors. These vectors are the following. $\vec{H}_0$ is a bias magnetic field, vector $\vec{p}$ shows a direction of wave propagation in a microwave waveguide, and $\vec{q}$ is a unit vector tangent to a curve depicting an angular squint. We can state that the three vectors $\vec{q}$, $\vec{p}$, $\vec{H}_0$ constitute the right-hand triple of vectors.

**C. A note on the balance of energy**

At the MDM resonance frequency, near a ferrite disk one observes the regions with positive and negative helicities and the near-field topology is distinguished by presence of the "helicity dipoles" ("ME dipoles"). At the MDM resonance frequency, a robust topological structure of a power-flow vortex near a ferrite disk is distorted (see Fig. 11). We showed that this distortion is in evident correlation with orientation of the ME dipole. We also showed that the far-field orbital angular momenta appear due to the near-field topology of the "helicity dipoles" ("ME dipoles"). It can be assumed that to create such "ME dipoles", some additional energy should be taken from RF sources. It means that appearance of the far-field orbital angular momenta should be accompanied with reduction of a gain in the radiation pattern. Such an estimation of balance of energy, made based on the numerical analysis, is illustrated in Fig. 20.

At the nonresonance frequency of MDM oscillations, a ferrite disk behaves as a small obstacle in a waveguide. Since the hole diameter is much less than a half wavelength of microwave radiation and the hole position does not lead cutting of electric currents in waveguide walls, we have negligibly small radiation power. We should not counting power lost due to joule heating in the feedline and reflections back down the feedline. As a results, at the nonresonance frequency of MDM oscillations, the antenna directivity and antenna gain are almost undistinguishable. This is shown in Fig. 20 (*a*).

The situation is completely different at the MDM resonance frequency. From the radiation patterns shown in Fig. 20 (*b*) it is evident that difference between the directivity and gain powers is about 19 dBi. This amount shows that the power loss due to joule heating in the feedline and reflections back down the feedline is about 98% from the input power in a waveguide. Based on an analysis of the reflection characteristic (shown in Fig. 2) and the transmission characteristic (shown in Fig. 20 (*c*)), one can easily estimate that the power reflection back down the feedline is not more than 40% from the input power. So, we are faced with the fact that a system has extremely big amount of the power losses due to joule heating. However, a simple analysis shows that the joule heating in the system is negligibly small. From a numerical reflection characteristics we can see that for the first "bright" peak – the peak $1'$ – the quality factor $Q$ is very big, about $10^4$ (from an experimental reflection characteristic we obtain that the quality factor for the peak $1'$ is about $2\times 10^3$). It becomes clear that for such a quantity $Q$, joule heating losses inside a ferrite disk are extremely small compared to the entire power losses in a MDM resonator. We also have negligibly small joule losses in waveguide walls. The conclusion is that the main losses take place due to absorption of energy necessary for creation of the "ME dipoles", which result in distortion of power-flow vortices near a ferrite disk and finally creation of the far-field orbital angular momenta in radiation characteristics.

### IV.  CONCLUSION

In terms of far-field scattering, the MDM particles have scattering patterns which are strongly not identical to the conventional electric and magnetic dipole (in general, multipole) structures. ME fields appear as a result of interaction of MDM oscillations in a ferrite particle with external EM fields. In the vicinity of a ferrite disk, at the MDM resonances one observes strongly localized areas of the electric and magnetic fields which are constituents of the ME field. While at the frequency beyond a MDM resonance one observes a current of a regular electric dipole, in a case of a MDM resonance, the surface electric current is a chiral current with an evident deviation of topological charges. The presence and positions of the in-plane regions with positive and negative ME helicity factor strongly influence on the distribution of power flows in the vicinity of the antenna radiation hole. From the pictures of ME-field areas on a vacuum plane above a radiation hole, one can see that there exist in-plane "ME dipoles" and that directions of these "ME dipoles" is strongly correlated with directions of the squints in radiation patterns. An approximate analysis of the balance of energy shows that the main losses in the antenna take place due to absorption energy necessary for creation of the "ME dipoles", which result in distortion of power-flow vortices near a ferrite disk and finally creation of the far-field orbital angular momenta in radiation characteristics. The approach showing that chiral-topology near fields originated from a small ferrite particle with MDM oscillations result in generating far-field OAM, bridges and combines the four concepts: (*a*) Electromagnetic (optical) chirality of the fields; (*b*) topological ME effects; (*c*) Fano resonances; (*d*) OAM antennas.

-------------------------------------------------------------------------------------------------------

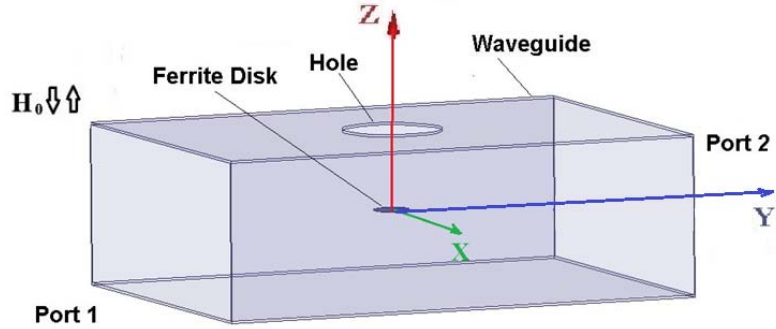

Fig. 1. MDM microwave antenna: a waveguide radiation structure with a hole in a wide wall and a thin-film ferrite disk as a basic building block. A ferrite disk is placed inside a $TE_{10}$-mode rectangular X-band waveguide symmetrically to its walls so that a disk axis is perpendicular to a wide wall of a waveguide. The hole diameter is much less than a half wavelength of microwave free-space radiation at the frequency regions used in the studies.

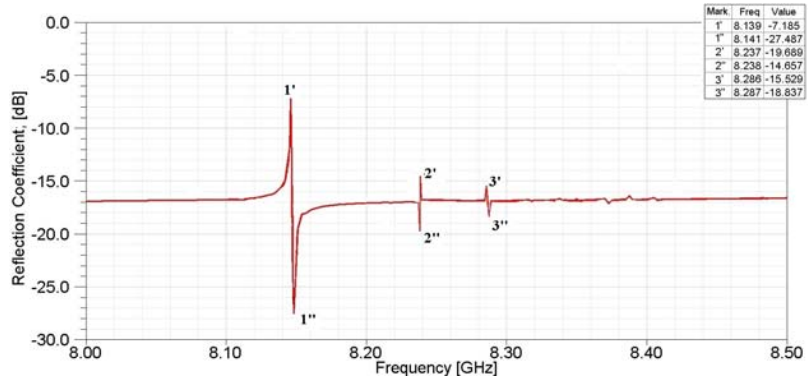

Fig. 2. Numerical reflection characteristic in a waveguide radiation structure.

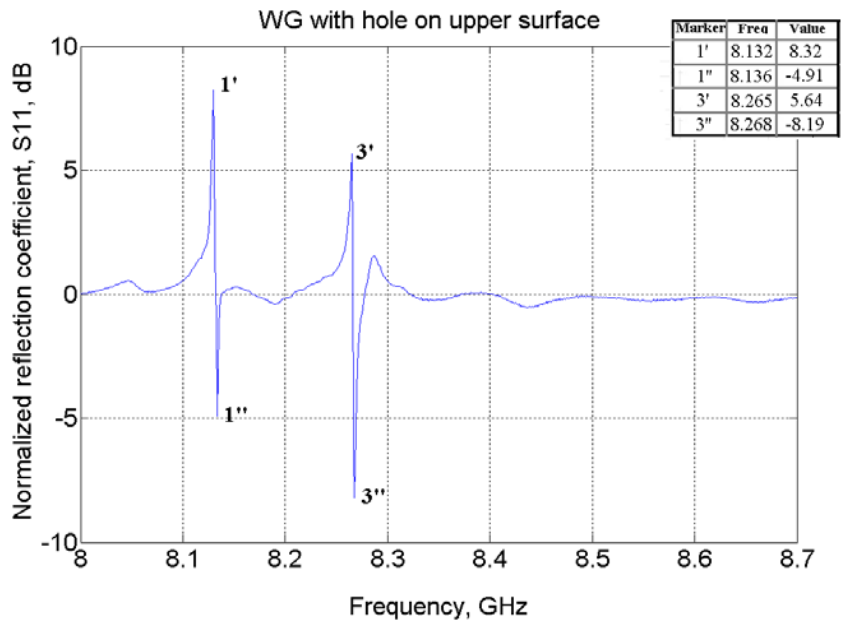

Fig. 3. Experimental reflection characteristic in a waveguide radiation structure.

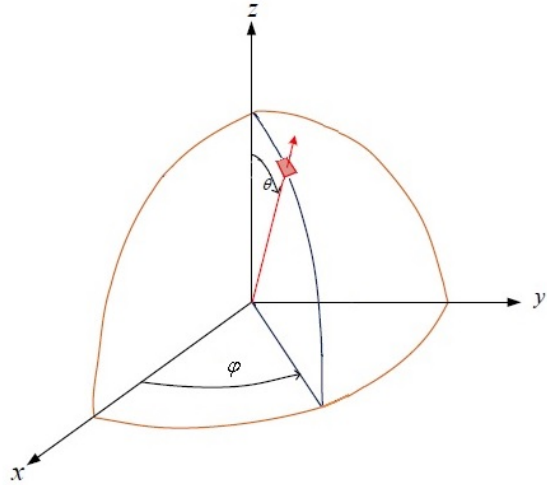

Fig. 4. Spherical coordinate system.

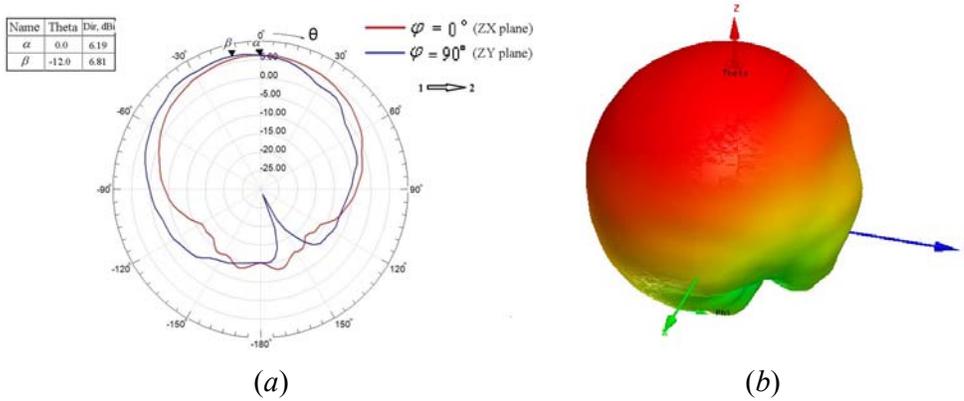

(*a*)  (*b*)

Fig. 5. Single-element radiation patterns at the nonresonance frequency 8.125GHz for two cut planes, $\varphi = 0°$ and $\varphi = 90°$. The wave in a waveguide propagates from port 1 to port 2 (**1 ⟹ 2**). (*a*) 2-D pattern; the squint is observed in the *zy* plane (marker $\beta$). No squint is observed in the *zx* plane (marker $\alpha$). (*b*) 3-D pattern. The radiation patterns are the same for two opposite directions of a bias magnetic field. The markers $\alpha$ and $\beta$ show maximal values of the directivity.

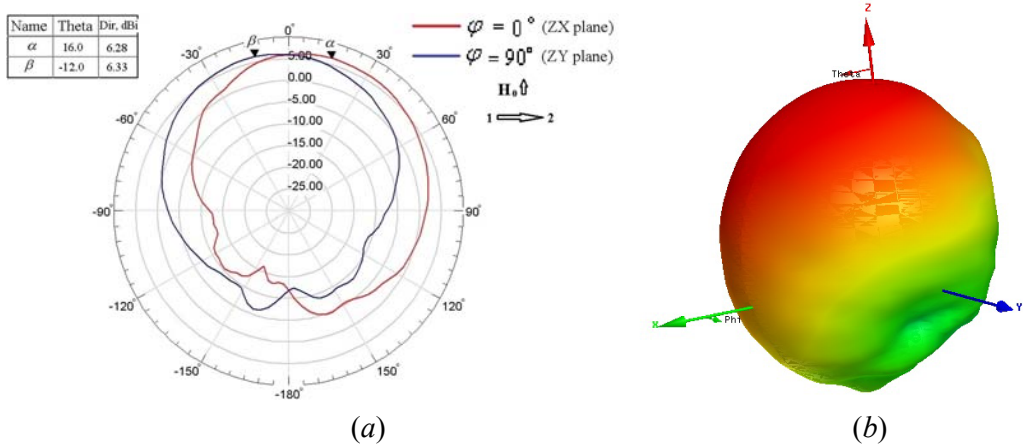

(a) (b)

Fig. 6. Single-element radiation patterns for two cut planes, $\varphi = 0°$ and $\varphi = 90°$, at the MDM-resonance frequency 8.139GHz. The wave in a waveguide propagates from port 1 to port 2. (*a*) 2-D pattern; (*b*) 3-D pattern. The squint in *zy* plane is the same as in Fig. 5. Upward direction of a bias field.

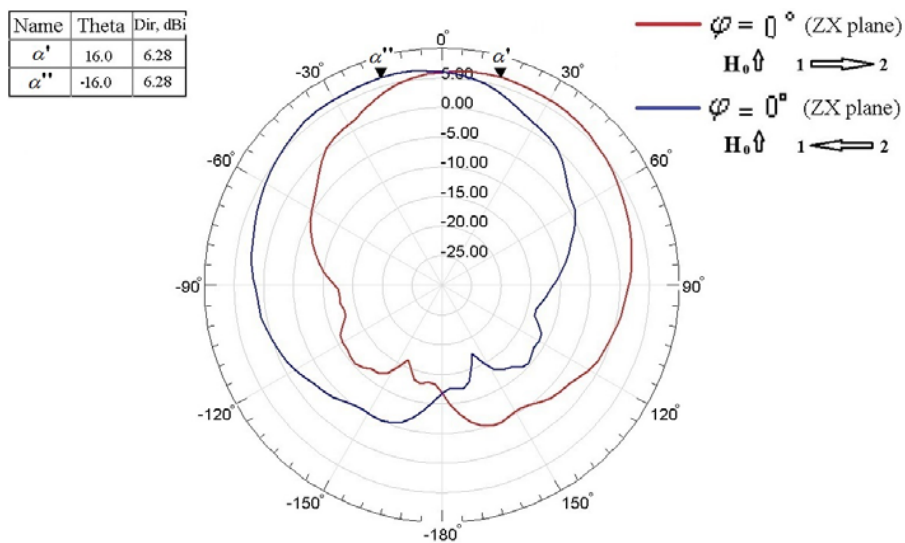

Fig. 7. Single-element radiation patterns for a cut plane $\varphi = 0°$ at the MDM-resonance frequency 8.139GHz and two opposite directions of the wave propagation in a waveguide. Red line – wave propagates from port 1 to port 2; blue line – wave propagates from port 2 to port 1. Upward direction of a bias field.

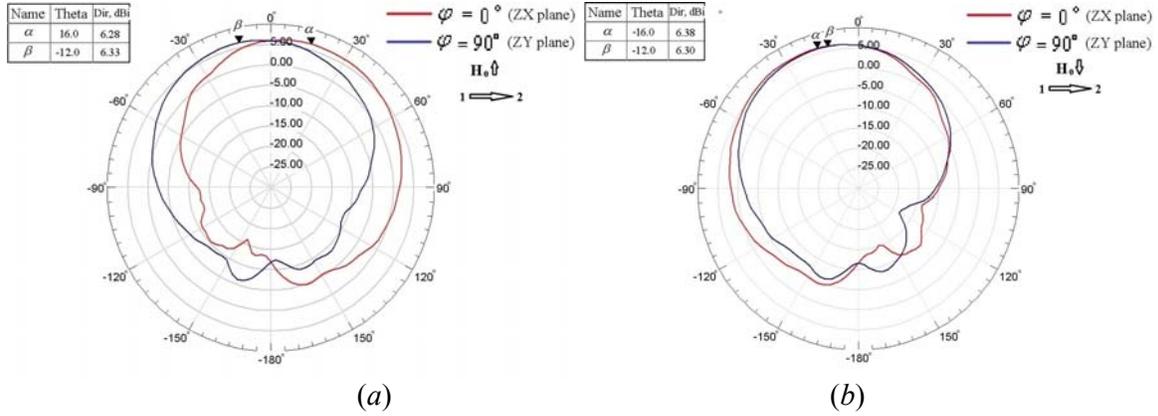

(*a*)                     (*b*)

Fig. 8. Resonance-frequency radiation patterns in the planes $\varphi = 0°$ and $\varphi = 90°$ for two opposite directions of the bias magnetic field and the same direction of the wave propagation in a waveguide (from port 1 to port 2). (*a*) Upward direction of a bias field, (*b*) downward direction of a bias field.

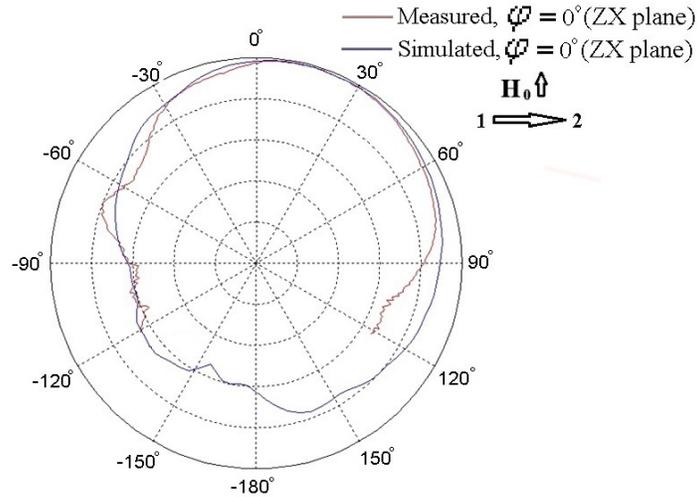

Fig. 9. Numerically simulated and experimentally measured radiation patterns for a cut plane $\varphi = 0°$ at the MDM-resonance frequency 8.139GHz. Wave propagates from port 1 to port 2. Upward direction of a bias field.

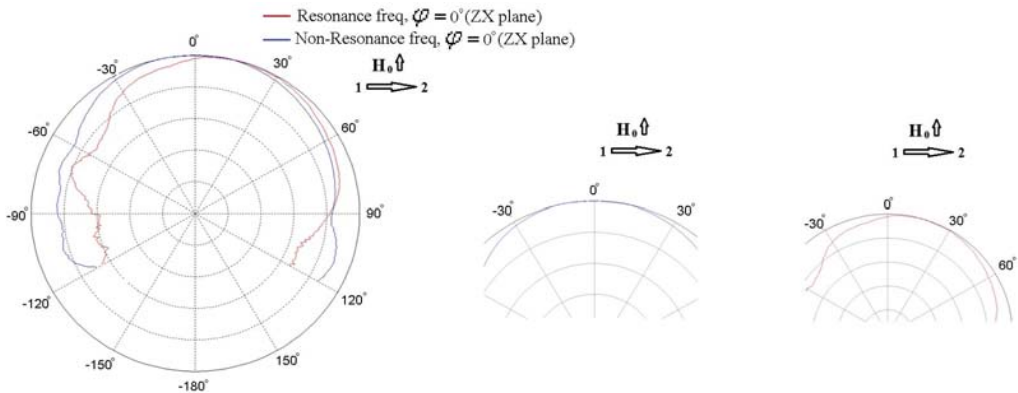

(*a*)                (*b*)                (*c*)

Fig. 10. Experimental radiation patterns for nonresonant (8.124GHz) and resonant (8.132GHz) frequencies. (*a*) A general picture of the normalized radiation patterns. (*b*) An enlarged picture of the radiation pattern for a nonresonant frequency. (*c*) An enlarged picture of the radiation pattern for a resonant frequency. The radiation patterns are measured in a *zx* plane ($\varphi=0°$). Wave propagates from port 1 to port 2 and a bias magnetic field is upwards directed.

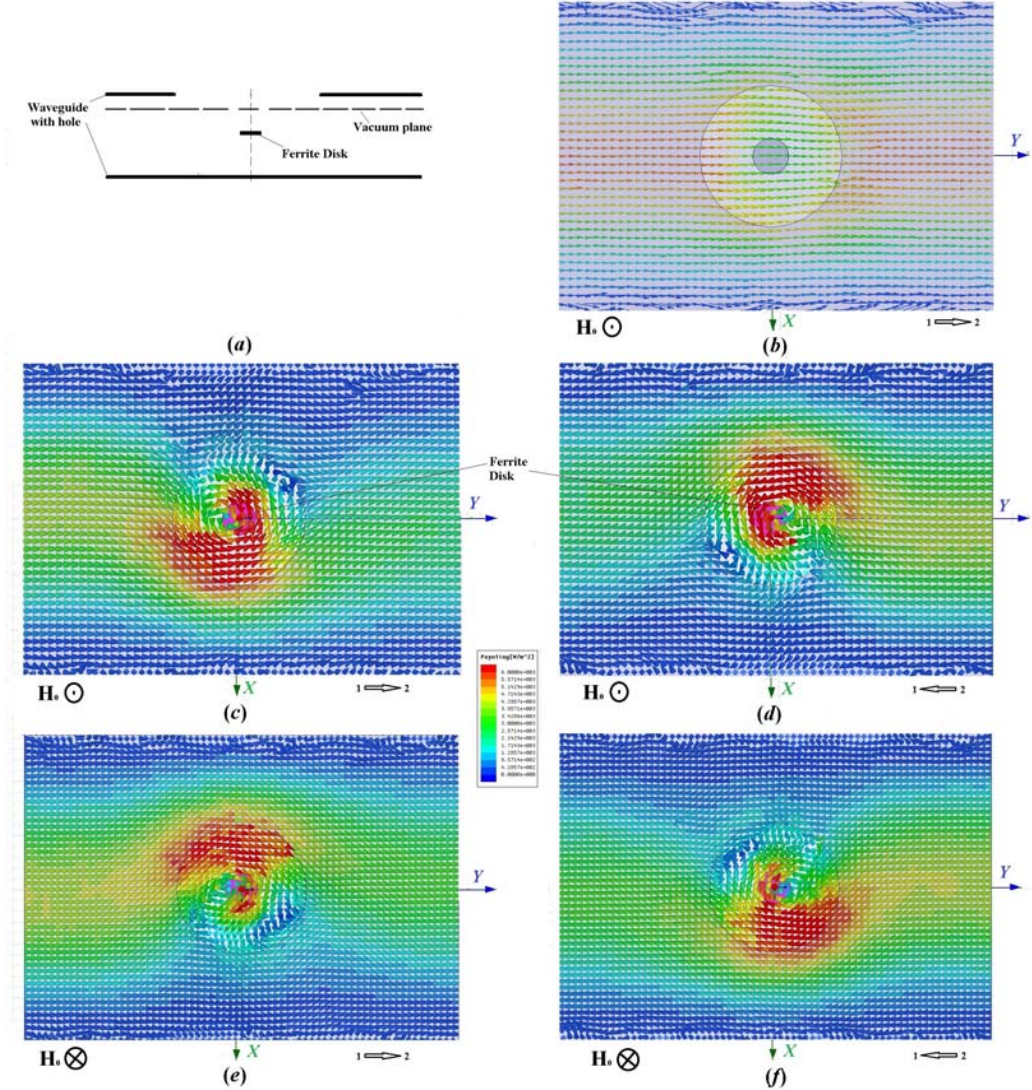

Fig. 11. Poynting-vector distributions on a vacuum plane below a radiation hole. (*a*) a schematic picture of a radiating structure with a vacuum plane; (*b*) wave propagation from port 1 to port 2 and upward direction of a bias field at a frequency beyond a MDM resonance; (*c*) wave propagation from port 1 to port 2 and upward direction of a bias field; (*d*) wave propagation from port 2 to port 1 and upward direction of a bias field; (*e*) wave propagation from port 1 to port 2 and downward direction of a bias field; (*f*) wave propagation from port 2 to port 1 and downward direction of a bias field.

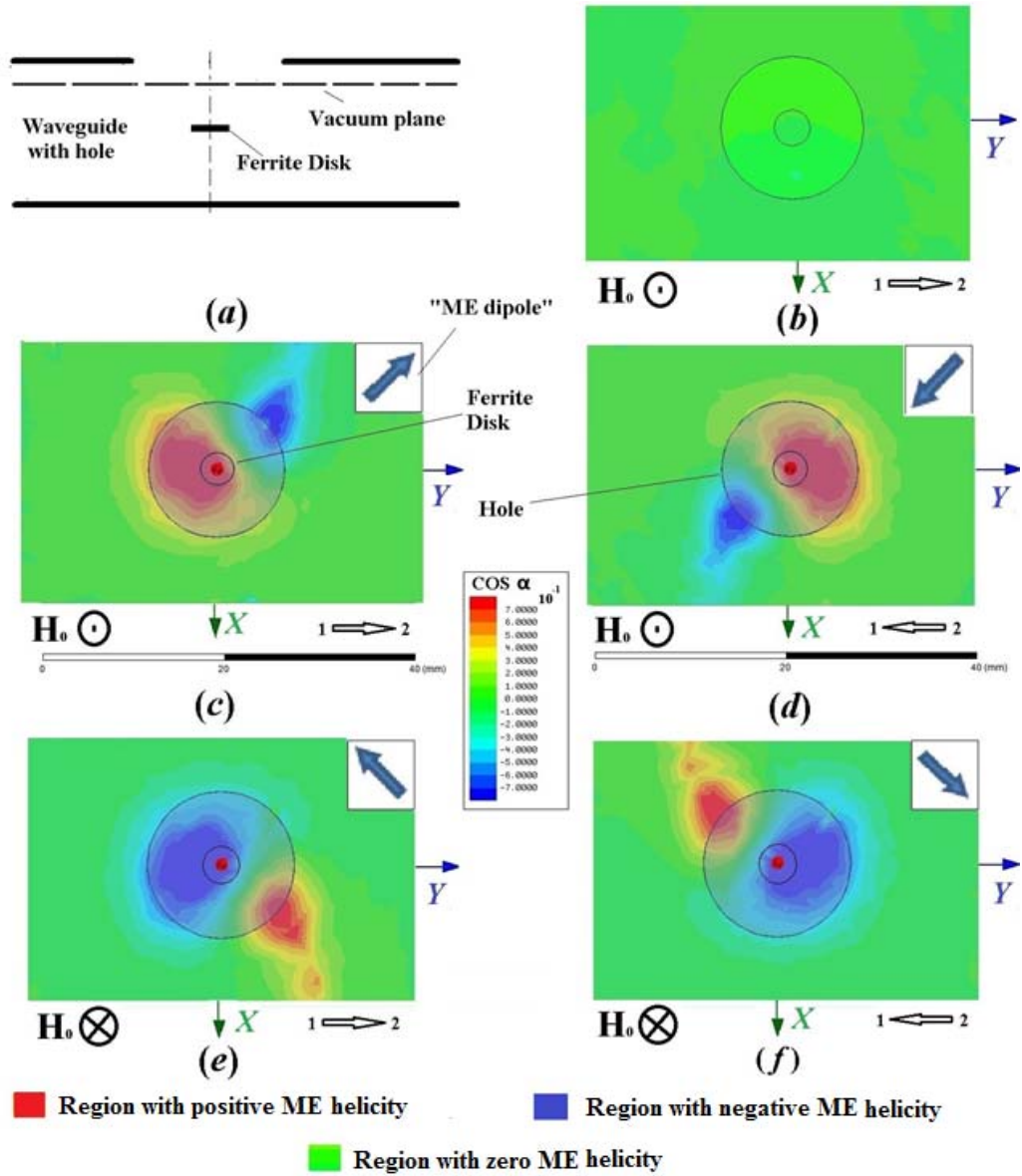

Fig. 12. ME-field areas on the *xy* plane on the vacuum plane below a radiation hole. (*a*) a schematic picture of a radiating structure with a vacuum plane; (*b*) wave propagation from port 1 to port 2 and upward direction of a bias field at a frequency beyond a MDM resonance; (*c*) wave propagation from port 1 to port 2 and upward direction of a bias field; (*d*) wave propagation from port 2 to port 1 and upward direction of a bias field; (*e*) wave propagation from port 1 to port 2 and downward direction of a bias field; (*f*) wave propagation from port 2 to port 1 and adownward direction of a bias field. The arrows in right-upper corners show directions of "ME dipoles": the directions from regions of the positive ME helicity to regions of the negative ME helicity.

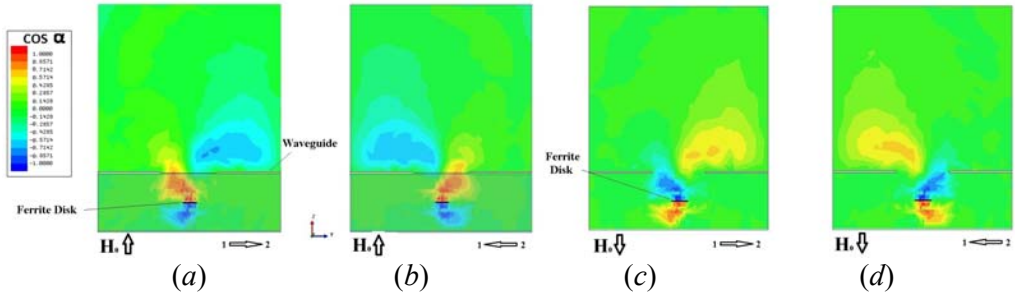

Fig. 13. ME-field areas on the cross-section $zy$ plane in near-field regions. (*a*) Wave propagation from port 1 to port 2 and upward direction of a bias field; (*b*) wave propagation from port 2 to port 1 and upward direction of a bias field; (*c*) wave propagation from port 1 to port 2 and downward direction of a bias field; (*d*) wave propagation from port 2 to port 1 and downward direction of a bias field.

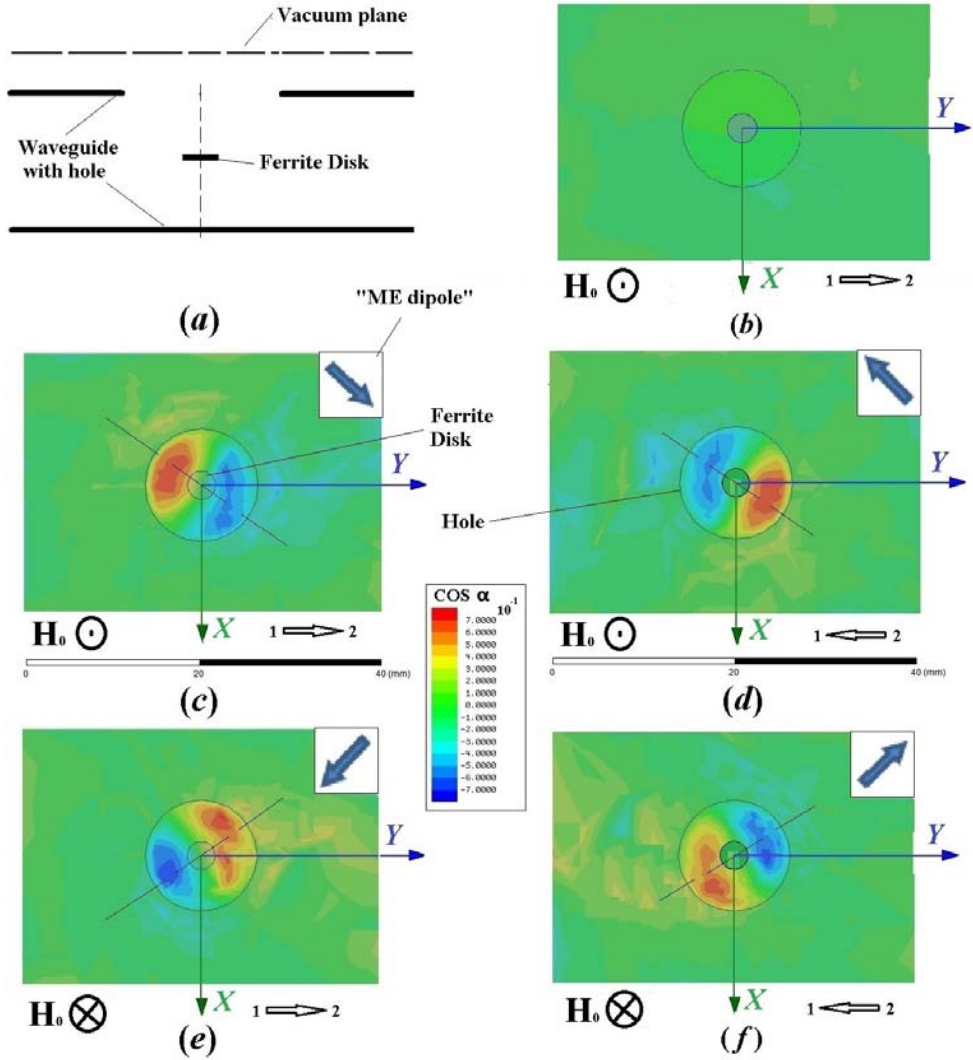

Fig. 14. ME-field areas on the $xy$ plane on the vacuum plane above a radiation hole. (*a*) a schematic picture of a radiating structure with a vacuum plane; (*b*) wave propagation from port 1 to port 2 and upward direction of a bias field at a frequency beyond a MDM resonance; (*c*)

wave propagation from port 1 to port 2 and upward direction of a bias field; (*d*) wave propagation from port 2 to port 1 and upward direction of a bias field; (*e*) wave propagation from port 1 to port 2 and downward direction of a bias field; (*f*) wave propagation from port 2 to port 1 and adownward direction of a bias field. The arrows in right-upper corners show directions of "ME dipoles": the directions from regions of the positive ME helicity to regions of the negative ME helicity.

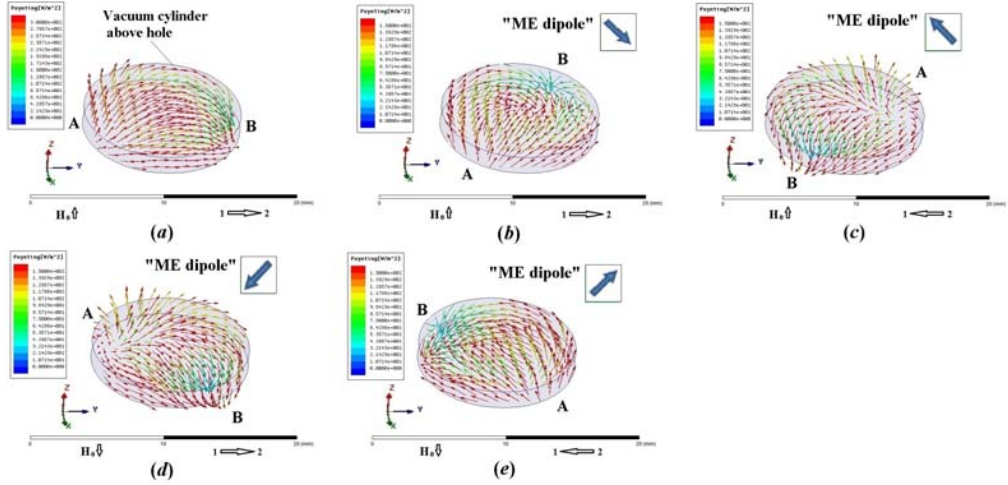

Fig. 15. Poynting-vector distributions in a vacuum cylinder above a radiation hole. (*a*) Wave propagation from port 1 to port 2 at a frequency beyond a MDM resonance; (*b*) MDM-resonance wave propagation from port 1 to port 2 and upward direction of a bias field; *c*) MDM-resonance wave propagation from port 2 to port 1 and upward direction of a bias field; (d) MDM-resonance wave propagation from port 1 to port 2 and downward direction of a bias field; (e) MDM-resonance wave propagation from port 2 to port 1 and downward direction of a bias field. The regions *A* and *B* conventionally designate the regions of positive and negative topological charges in the power-flow-density distributions.

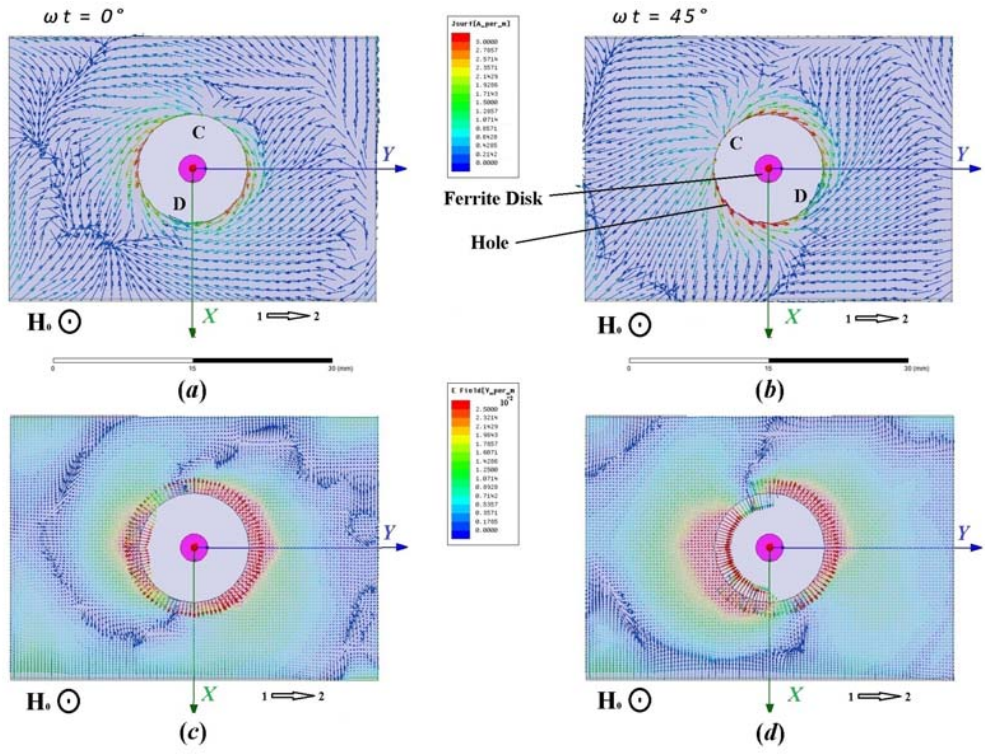

Fig. 16. Electric currents (*a*), (*b*) and fields (*c*), (*d*) on the external metal surface of a waveguide at the frequency MDM resonance, 8.139GHz. (*a*), (*c*) Time phase (*a*), (*b*) $\omega t = 0°$; (*b*), (*d*) time phase $\omega t = 45°$. MDM-resonance wave propagation from port 1 to port 2 and upward direction of a bias magnetic field;

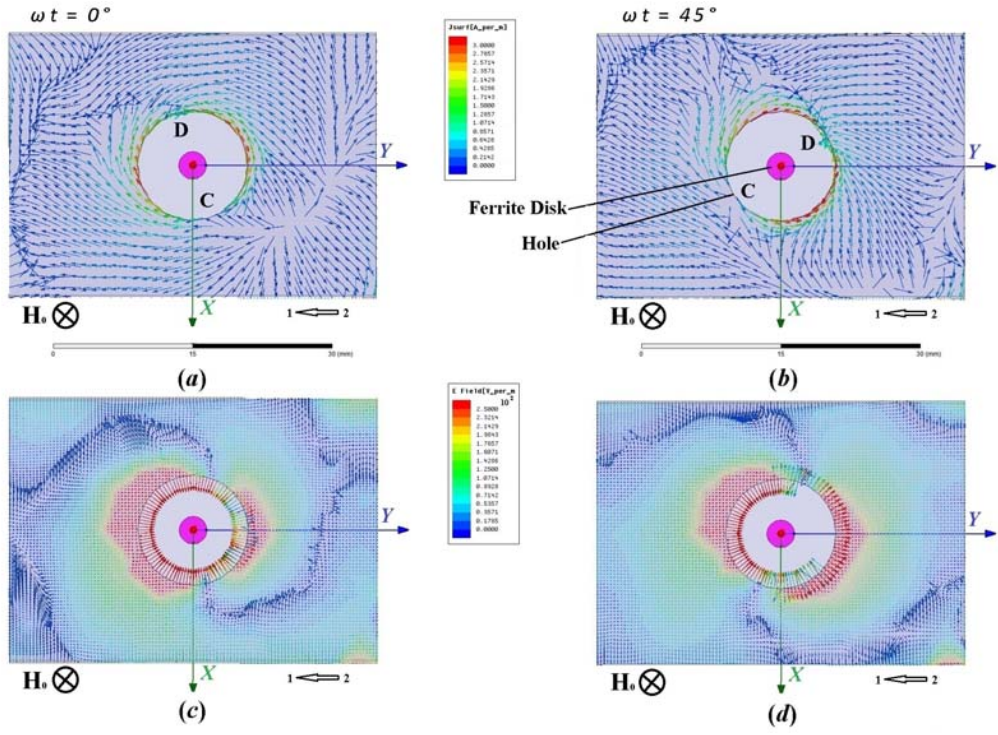

Fig. 17. Electric currents (*a*), (*b*) and fields (*c*), (*d*) on the external metal surface of a waveguide at the frequency MDM resonance, 8.139GHz. (*a*), (*c*) Time phase (*a*), (*b*) $\omega t = 0°$; (*b*), (*d*) time phase $\omega t = 45°$. MDM-resonance wave propagation from port 2 to port 1 and downward direction of a bias magnetic field.

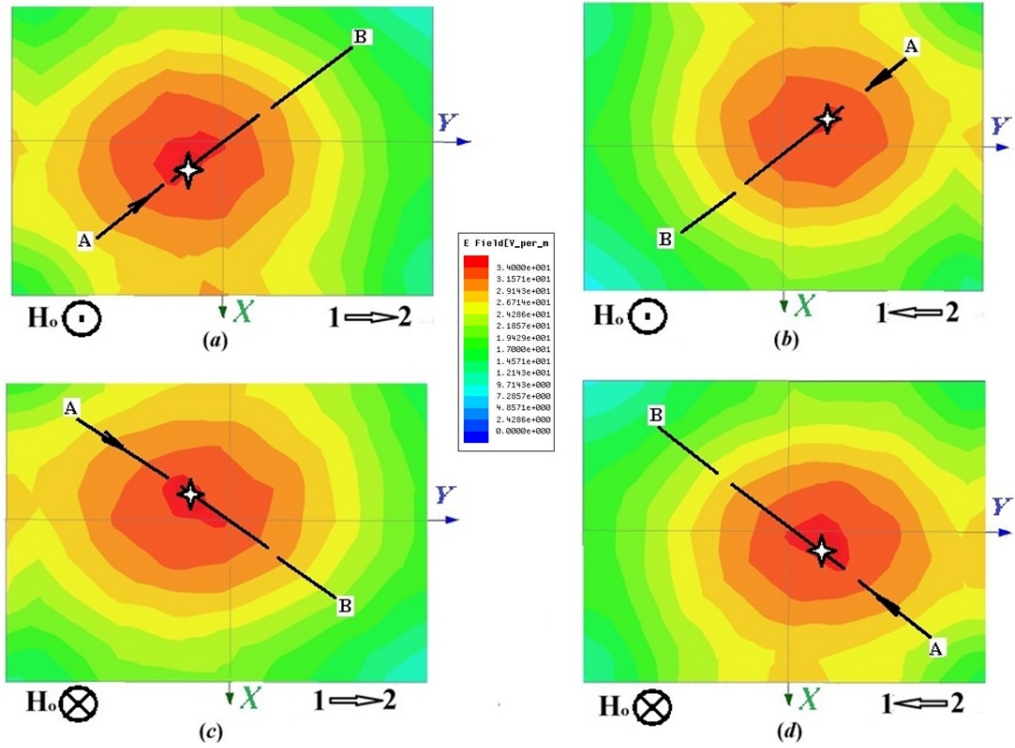

Fig. 18. Intensity of an electric field on the *xy* vacuum plane situated at distance about $3\lambda_0$ above a radiation hole. The figure shows time averaged directions of the squints for different directions of a bias field and wave propagation in a waveguide. The *AB* lines connect the regions of positive and negative topological charges of the power-flow-density distributions in the near-field region. Small stars, show the places of maximal intensities of the electric field on *AB* lines. The squint positions are in strong correlation with the directions and orientations of the *AB*-line vectors.

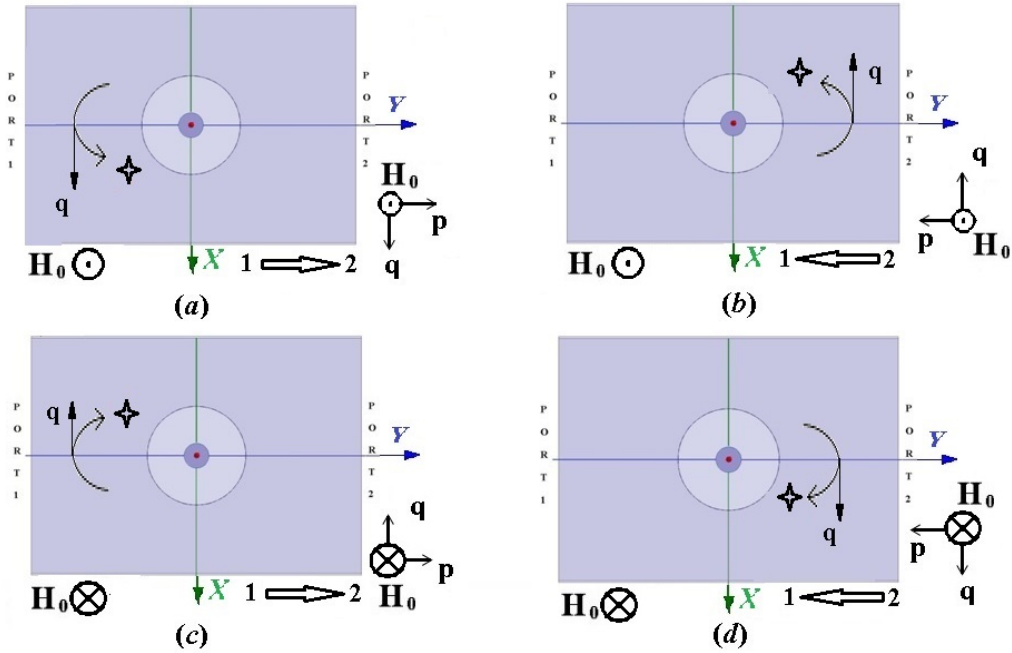

Fig. 19. Distribution of maximal intensities of an electric field on the *xy* vacuum plane in correlation with directions of the three vectors: a direction of a bias magnetic field, a direction of wave propagation in a waveguide, and a unit vector tangent to a curve depicting an angular squint. The three vectors $\vec{q}$, $\vec{p}$, $\vec{H}_0$ constitute the right-hand triple of vectors.

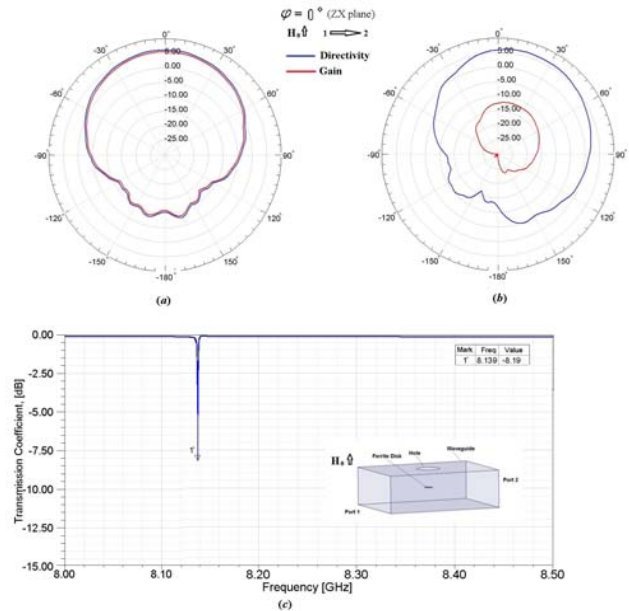

Fig. 20. Balance of energy. (*a*) The directivity and gain at the nonresonance frequency 8.125GHz. (*b*) The directivity and gain at the MDM-resonance frequency 8.139GHz, when the orbital angular momentum is observed. (*c*) The waveguide transmission coefficient characteristics.